\documentclass[aps,pre,twocolumn,showpacs]{revtex4}
\usepackage{color,amsmath,graphicx}
\newcommand{\vtr}[1]{\mbox{\boldmath $#1$}}
\begin{document}

\title{%
Reformulation of the Stochastic Potential Switching Algorithm and 
a Generalized Fourtuin-Kasteleyn Representation
}

\author{Munetaka Sasaki}
\affiliation{Department of Applied Physics, Tohoku University, Sendai, 980-8579}

\date{\today}

\begin{abstract}
A new formulation of the stochastic potential switching algorithm is presented. 
This reformulation naturally leads us to a generalized Fourtuin-Kasteleyn 
representation of the partition function $Z$. 
A formula for internal energy $E$ and that of heat capacity $C$ are derived 
from derivatives of the partition function. We also derive a formula for the 
exchange probability in the replica exchange Monte Carlo method. By combining the formulae
with the Stochastic Cutoff method, we can greatly reduce the computational time 
to perform internal energy and heat capacity measurements and the replica exchange 
Monte Carlo method in long-range interacting systems. Numerical simulations in three 
dimensional magnetic dipolar systems show the validity of the method. 
\end{abstract}
\maketitle

\section{Introduction}
One of the most challenging subjects in the field of 
computational physics is to develop efficient methods 
for long-range interacting systems. The difficulty of 
long-range interacting systems comes from the fact 
that the number of interactions rapidly increases 
with increasing the number of elements $N$ of the system. 
For example, in the case of systems with pairwise interactions, 
the number of interactions is proportional to $N^2$. 
Therefore, if one uses a naive simulation method in such systems, 
the computational time per one step rapidly increases in proportion 
to $N^2$, which is quite contrast to the case of short-range 
interacting systems in which the computational time increases
in proportion to $N$. In order to overcome the difficulty, 
many simulation methods have been proposed until
now~\cite{Appel85,Barnes86,Greengard88,Carrier88,Saito92,Ding92,LuijtenBlote95,Sasaki96,
Hetenyi02, Bernacki04}. 

Recently, the author and Matsubara have developed an efficient 
Monte Carlo (MC) method called Stochastic CutOff (SCO) method 
for long-range interacting systems~\cite{SasakiMatsubara08}. 
The basic idea of the method is to switch long-range 
interactions $V_{ij}$ stochastically 
to either zero or a pseudointeraction ${\bar V}_{ij}$ 
with the detailed balance condition satisfied. 
Interactions are switched by using the Stochastic Potential Switching (SPS) 
algorithm~\cite{Mak05,MakSharma07}. Because most of the distant and weak 
interactions are eliminated by being switched to zero, the SCO method 
greatly reduces the number of interactions and computational time 
in long-range interacting systems~\cite{TimeReduction}. 
Furthermore, this method does not involve any approximation 
because the detailed balance condition is satisfied strictly. 
Fukui and Todo have recently developed an efficient MC 
method based on similar strategy by use of different 
pseudointeractions and different way of switching 
interactions~\cite{FukuiTodo09}. 

In the present work, we reformulate the SPS algorithm which is 
used in the SCO method. This reformulation gives us a generalized version of 
the Fourtuin-Kasteleyn representation of the partition function 
in the Ising ferromagnetic model~\cite{KasteleynFortuin69,FortuinKasteleyn72}. 
This representation of the partition function is used 
to derive new formulae for internal energy and heat capacity measurements. 
Since these formulae consist of only terms which survive as $\bar V_{ij}$ 
in the above-mentioned potential switching process, the computational time for the measurements 
are reduced greatly. We also derive an formula which reduces the computational 
time to estimate exchange probability in the replica exchange MC method~\cite{HukushimaNemoto96}. 
In order to verify the formulae, we perform some MC simulations 
in three dimensional magnetic dipolar systems. The results clearly show 
the validity of the formulae.

The organization of the paper is as follows. In \S\ref{sec:reformulation}, 
we reformulate the stochastic potential switching algorithm. A generalized 
Fourtuin-Kasteleyn representation of the partition function is presented in this section. 
In \S\ref{sec:formulaA} and \S\ref{sec:formulaB}, we derive formulae for internal energy 
and heat capacity measurements and a formula for the replica exchange MC method, 
respectively. The validity of these formulae is confirmed numerically in 
\S\ref{sec:NumericalTests}. Section \ref{sec:summary} is devoted to a summary 
and discussions.

\section{Reformulation of the SPS Algorithm}
\label{sec:reformulation} 
Before reformulating the SPS algorithm, we briefly explain 
the SPS algorithm~\cite{Mak05,MakSharma07}. 
We hereafter consider a system with pairwise 
long-range interactions described by the Hamiltonian
\begin{equation}
{\cal H}=\sum\nolimits_{i<j}V_{ij}(\vtr{S}_i,\vtr{S}_j),
\label{eqn:GH}
\end{equation} 
where $\vtr{S}_i$ is a variable associated with the $i$-th element 
of the system. In this algorithm, $V_{ij}$ is stochastically switched 
to either $\tilde{V}_{ij}$ or $\bar{V}_{ij}$ 
with a probability of $P_{ij}$ or $1-P_{ij}$, respectively. 
The probability $P_{ij}$ is 
\begin{equation}
P_{ij}(\vtr{S}_i,\vtr{S}_j)=
\exp[\beta (\Delta V_{ij}(\vtr{S}_i,\vtr{S}_j)-\Delta V_{ij}^*)],
\label{eqn:Sprob}
\end{equation}
where $\beta$ is the inverse temperature, 
\begin{equation}
\Delta V_{ij}(\vtr{S}_i,\vtr{S}_j)\equiv V_{ij}(\vtr{S}_i,\vtr{S}_j)-
\tilde V_{ij} (\vtr{S}_i,\vtr{S}_j),
\label{eqn:DefofDeltaV}
\end{equation} 
and $\Delta V_{ij}^*$ is a constant equal to (or greater than) 
the maximum value of $\Delta V_{ij}(\vtr{S}_i,\vtr{S}_j)$ over all 
$\vtr{S}_i$ and $\vtr{S}_j$. We can choose the potential 
$\tilde{V}_{ij}$ {\it arbitrarily}. 
On the other hand, using $P_{ij}(\vtr{S}_i,\vtr{S}_j)$, 
the potential $\bar{V}_{ij}$ is given as 
\begin{equation}
\bar{V}_{ij}(\vtr{S}_i,\vtr{S}_j)=
V_{ij}(\vtr{S}_i,\vtr{S}_j)
-\beta^{-1}\log[1-P_{ij}(\vtr{S}_i,\vtr{S}_j)].
\label{eqn:Ppseudo}
\end{equation}
With this potential switching process, the algorithm proceeds as follows:
\begin{itemize}
\item[(A)] Potentials $V_{ij}$ are switched 
to either $\tilde{V}_{ij}$ or $\bar{V}_{ij}$ with the probability of 
$P_{ij}$ or $1-P_{ij}$, respectively.
\item[(B)] A standard MC simulation is performed with the 
switched Hamiltonian ${\cal H}'$ expressed as 
\begin{equation}
{\cal H}'=\sum\nolimits'\tilde{V}_{ij}(\vtr{S}_i,\vtr{S}_j)+
\sum\nolimits''\bar{V}_{ij}(\vtr{S}_i,\vtr{S}_j),
\label{eqn:Spotential}
\end{equation}
where $\sum'$ runs over all the potentials switched to $\tilde{V}$ 
and $\sum''$ runs over those switched to $\bar{V}$. 
The potential is fixed during the simulation. 
\item[(C)] Return to (A).
\end{itemize}
It is shown that this MC procedure strictly satisfies 
the detailed balance condition with respect to the original Hamiltonian 
of eq.~(\ref{eqn:GH}). 

In the SCO method, ${\tilde V}_{ij}$ is set to zero to reduce 
the computational time of ${\cal H}'$ defined by eq.~(\ref{eqn:Spotential}). 
Furthermore, the use of an efficient method enables us to reduce the 
computational time of the potential switching procedure (A) greatly
(see ref.~\cite{SasakiMatsubara08} for details). 

In order to reformulate the SPS algorithm, 
we first introduce a graph variable $g_{ij}$ defined by
\begin{equation}
g_{ij}=\left\{
\begin{array}{cc}
0 & \textrm{if $V_{ij}$ is switched to ${\tilde V}_{ij}$},\\
1 & \textrm{if $V_{ij}$ is switched to ${\bar V}_{ij}$}.\\
\end{array}
\label{eqn:DefGij}
\right.
\end{equation}
We next introduce a weight $\omega(\vtr{S}_i,\vtr{S}_j;g_{ij})$ 
defined by 
\begin{equation}
\omega_{ij}(\vtr{S}_i,\vtr{S}_j;g_{ij})=\left\{
\begin{array}{cc}
{\rm e}^{-\beta \{{\tilde V}_{ij}(\vtr{S}_i,\vtr{S}_j)+\Delta V_{ij}^* \}}
& (g_{ij}=0),\vspace{1mm}\\
{\rm e}^{-\beta {\bar V}_{ij}(\vtr{S}_i,\vtr{S}_j)} & (g_{ij}=1).
\label{eqn:defomega}
\end{array}
\right.\\
\end{equation}
This weight is analogous to the weight introduced by 
Edwards and Sokal~\cite{EdowardsSokal88}. 
We hereafter show that the SPS algorithm is a MC 
method which realizes equilibrium distribution defined by
\begin{equation}
P_{\rm SPS}(\{\vtr{S}_{i}\},\{g_{ij}\})
\equiv Z_{\rm SPS}^{-1}\prod\nolimits_{i<j} \omega_{ij}(\vtr{S}_i,\vtr{S}_j;g_{ij}),
\label{eqn:ProbSPS}
\end{equation}
where
\begin{equation}
Z_{\rm SPS}\equiv{\rm Tr}_{\{\vtr{S}_i\},\{g_{ij}\}}
\prod\nolimits_{i<j} \omega_{ij}(\vtr{S}_i,\vtr{S}_j;g_{ij}).
\label{eqn:PartitionSPS}
\end{equation}
As it is readily derived from eqs.~(\ref{eqn:Sprob}), (\ref{eqn:DefofDeltaV}), 
(\ref{eqn:Ppseudo}), and (\ref{eqn:defomega}), the weight 
$\omega_{ij}(\vtr{S}_i,\vtr{S}_j;g_{ij})$ has the following property:
\begin{equation}
{\rm Tr}_{g_{ij}=0,1}\omega_{ij}(\vtr{S}_i,\vtr{S}_j;g_{ij})
=\exp[-\beta V_{ij}(\vtr{S}_i,\vtr{S}_j)].
\label{eqn:TraceOmega}
\end{equation}
This equation naturally leads us to the following new representation of 
the partition function
\begin{equation}
Z(\beta)=Z_{\rm SPS}(\beta)={\rm Tr}_{\{\vtr{S}_i\}, \{g_{ij}\}} 
\prod\nolimits_{i<j} \omega_{ij}(\vtr{S}_i,\vtr{S}_j;g_{ij}).
\label{eqn:GeneralizedFK}
\end{equation}
We also find that the probability that some configuration $\{\vtr{S}_i\}$ 
is realized in the SPS algorithm is given as
\begin{eqnarray}
P(\{\vtr{S}_i\})={\rm Tr}_{\{g_{ij}\}} 
P_{\rm SPS} (\{\vtr{S}_{i}\},\{g_{ij}\})
=P_{\rm B}(\{\vtr{S}_i\}),
\label{eqn:RBandSPS}
\end{eqnarray}
where $P_{\rm B}$ is the Boltzmann distribution defined as
\begin{equation}
P_{\rm B}(\{\vtr{S}_i\})=Z(\beta)^{-1}
\exp\left[-\beta\sum\nolimits_{i<j}V_{ij}(\vtr{S}_i,\vtr{S}_j)\right].
\label{eqn:Bdistribution}
\end{equation}
This is the reason why the Boltzmann sampling is achieved by the SPS algorithm.

We next show that the equilibrium distribution of the SPS algorithm 
is given by eq.~(\ref{eqn:ProbSPS}). Let us denote the transition 
probability in the step (A) of the SPS algorithm 
as $W_{\rm A}(\{g_{ij}\}\rightarrow\{g_{ij}'\} | \{\vtr{S}_i\})$
and that in the step (B) as 
$W_{\rm B}(\{\vtr{S}_i\}\rightarrow\{\vtr{S}_i'\} | \{g_{ij}\})$. 
It should be noted that the procedure in the step (A) updates 
the graph variables $\{g_{ij}\}$ with fixing 
the configuration variables $\{\vtr{S}_i\}$, 
and that in the step (B) updates $\{\vtr{S}_i\}$ with fixing $\{g_{ij}\}$. 
In the following, we will show that the two transition probabilities 
satisfy the detailed balance conditions
\begin{eqnarray}
&&\hspace*{-8mm}P_{\rm SPS}(\{\vtr{S}_{i}\},\{g_{ij}\})
W_{\rm A}(\{g_{ij}\}\rightarrow\{g_{ij}'\} | \{\vtr{S}_i\})\nonumber\\
&&\hspace*{-8mm}=P_{\rm SPS}(\{\vtr{S}_{i}\},\{g_{ij}'\})
W_{\rm A}(\{g_{ij}'\}\rightarrow\{g_{ij}\} | \{\vtr{S}_i\}),
\label{eqn:DconditionA}
\end{eqnarray}
\begin{eqnarray}
&&\hspace*{-8mm}P_{\rm SPS}(\{\vtr{S}_{i}\},\{g_{ij}\})
W_{\rm B}(\{\vtr{S}_i\}\rightarrow\{\vtr{S}_i'\} | \{g_{ij}\})\nonumber\\
&&\hspace*{-8mm}=P_{\rm SPS}(\{\vtr{S}_{i}'\},\{g_{ij}\})
W_{\rm B}(\{\vtr{S}_i'\}\rightarrow\{\vtr{S}_i\} | \{g_{ij}\}).
\label{eqn:DconditionB}
\end{eqnarray}
These two equations clearly show that the equilibrium distribution 
of the SPS algorithm is $P_{\rm SPS}(\{\vtr{S}_{i}\},\{g_{ij}\})$. 

We start from the proof of eq.~(\ref{eqn:DconditionA}). 
It can be easily seen from the procedure in the step (A) that 
\begin{eqnarray}
&&\hspace*{-5mm}W_{\rm A}(\{g_{ij}\}\rightarrow\{g_{ij}'\} | \{\vtr{S}_i\})
\nonumber\\
&&\hspace*{-5mm}=\prod\nolimits^{(0)'} P_{ij}(\vtr{S}_i,\vtr{S}_j)
\prod\nolimits^{(1)'} [1-P_{ij}(\vtr{S}_i,\vtr{S}_j)],
\label{eqn:WAfirst}
\end{eqnarray}
where the product $\prod\nolimits^{(0)'}$ runs over 
all the pairs with $g_{ij}'=0$ and $\prod\nolimits^{(1)'}$ 
runs over those with $g_{ij}'=1$. It should be noted that 
$W_{\rm A}$ does not depend on $\{g_{ij}\}$. 
To rewrite the right hand side of the above equation, 
we utilize the following two equations: 
Firstly, it is found from eqs.~(\ref{eqn:Sprob}), 
(\ref{eqn:DefofDeltaV}), and (\ref{eqn:defomega}) that 
\begin{equation}
P_{ij}(\vtr{S}_i,\vtr{S}_j)=\omega_{ij}(\vtr{S}_i,\vtr{S}_j;g_{ij}'=0)
{\rm e}^{\beta V_{ij}(\vtr{S}_i,\vtr{S}_j)}.
\label{eqn:RewritePij}
\end{equation}
Secondly, we can rewrite $1-P_{ij}(\vtr{S}_i,\vtr{S}_j)$ as
\begin{eqnarray}
&&1-P_{ij}(\vtr{S}_i,\vtr{S}_j)\nonumber\\
&&=\left\{{\rm e}^{-\beta V_{ij}(\vtr{S}_i,\vtr{S}_j)}
-\omega_{ij}(\vtr{S}_i,\vtr{S}_j;g_{ij}'=0)\right\}
{\rm e}^{\beta V_{ij}(\vtr{S}_i,\vtr{S}_j)}\nonumber\\
&&=\omega_{ij}(\vtr{S}_i,\vtr{S}_j;g_{ij}'=1)
{\rm e}^{\beta V_{ij}(\vtr{S}_i,\vtr{S}_j)},
\end{eqnarray}
where we have used eqs.~(\ref{eqn:TraceOmega}) and (\ref{eqn:RewritePij}). 
By substituting these two equations into eq.~(\ref{eqn:WAfirst}), 
we obtain 
\begin{eqnarray}
&&\hspace*{-7mm}W_{\rm A}(\{g_{ij}\}\rightarrow\{g_{ij}'\} | \{\vtr{S}_i\})\nonumber\\
&&\hspace*{-7mm}=\prod\nolimits_{i<j}\omega_{ij}(\vtr{S}_i,\vtr{S}_j;g_{ij}')
{\rm e}^{\beta V_{ij}(\vtr{S}_i,\vtr{S}_j)}\nonumber \\
&&\hspace*{-7mm}=Z_{\rm SPS} P_{\rm SPS}(\{\vtr{S}_{i}\},\{g_{ij}'\})\prod\nolimits_{i<j}
{\rm e}^{\beta V_{ij}(\vtr{S}_i,\vtr{S}_j)}.
\end{eqnarray}
This equation shows that $W_{\rm A}$ satisfies 
the detailed balance condition (\ref{eqn:DconditionA}). 

Our second task is to prove eq.~(\ref{eqn:DconditionB}). 
Since we perform a MC simulation with the switched 
Hamiltonian ${\cal H}'$, the transition probability $W_{\rm B}$ 
satisfies the detailed balance condition
\begin{eqnarray}
&&\hspace*{-8mm}{\rm e}^{-\beta\left\{\sum^{(0)} {\tilde V}_{ij}(\vtr{S}_i,\vtr{S}_j)
+\sum^{(1)} {\bar V}_{ij}(\vtr{S}_i,\vtr{S}_j)\right\}}\nonumber\\
&&\hspace*{15mm}\times W_{\rm B}(\{\vtr{S}_i\}\rightarrow\{\vtr{S}_i'\} | \{g_{ij}\})\nonumber\\
&&\hspace*{-8mm}={\rm e}^{-\beta\left\{\sum^{(0)} {\tilde V}_{ij}(\vtr{S}_i',\vtr{S}_j')
+\sum^{(1)} {\bar V}_{ij}(\vtr{S}_i',\vtr{S}_j')\right\}}\nonumber\\
&&\hspace*{15mm}\times W_{\rm B}(\{\vtr{S}_i'\}\rightarrow\{\vtr{S}_i\} | \{g_{ij}\}),
\end{eqnarray}
where the sum $\sum\nolimits^{(0)}$ runs over 
all the pairs with $g_{ij}=0$ and $\sum\nolimits^{(1)}$ 
runs over those with $g_{ij}=1$. By multiplying 
${\rm e}^{-\beta \sum^{(1)}\Delta V_{ij}^*}$ to 
the both hand sides of the equation, we obtain
\begin{eqnarray}
&&\hspace*{-12mm}
W_{\rm B}(\{\vtr{S}_i\}\rightarrow\{\vtr{S}_i'\} | \{g_{ij}\})
\prod\nolimits_{i<j}\omega_{ij}(\vtr{S}_i,\vtr{S}_j;g_{ij})\nonumber\\
&&\hspace*{-12mm}
=W_{\rm B}(\{\vtr{S}_i'\}\rightarrow\{\vtr{S}_i\} | \{g_{ij}\})
\prod\nolimits_{i<j}\omega_{ij}(\vtr{S}_i',\vtr{S}_j';g_{ij}), 
\end{eqnarray}
where we have used eq.~(\ref{eqn:defomega}). 
It is clear from eq.~(\ref{eqn:ProbSPS}) that 
this equation is equivalent to the detailed balance 
condition~(\ref{eqn:DconditionB}).

We next turn to the new representation of the partition
function, {\it i.e.}, eq.~(\ref{eqn:GeneralizedFK}). 
This is a generalization of the Fourtuin-Kasteleyn 
representation of the partition function in the 
Ising ferromagnetic model~\cite{KasteleynFortuin69,FortuinKasteleyn72}. 
In fact, it is shown in the appendix~A 
that the original Fourtuin-Kasteleyn representation 
is derived from eq.~(\ref{eqn:GeneralizedFK}). 
This representation is more comprehensive than the original 
one in the following two points:
\begin{itemize}
\item[1)] In the new representation, potential 
${\tilde V}_{ij}$ can be chosen arbitrarily. 
On the other hand, the original Fourtuin-Kasteleyn representation
corresponds to a special case in which ${\tilde V}_{ij}$ is zero. 
\item[2)] The new representation is valid no matter whether 
the configuration variables $\{ \vtr{S}_i \}$ are discrete or continuous. 
This is contrast to the original representation which is derived 
for systems with discrete variables. 
\end{itemize}
This generalized Fourtuin-Kasteleyn representation will be used in the 
next section to derive formulae for internal energy and heat capacity 
measurements.

\section{Formulae for Internal Energy and Heat Capacity}
\label{sec:formulaA}
In order to derive formulae for internal energy $\langle E \rangle $ and 
heat capacity $\langle C \rangle$, we use the following thermodynamic relations:
\begin{equation}
\langle E \rangle =-Z^{-1} \frac{\partial Z}{\partial \beta},
\label{eqn:InternalEnergy0}
\end{equation}
\begin{equation}
\langle C\rangle = k_{\rm B}\beta^{2}\left(
Z^{-1}\frac{\partial^2 Z}{\partial \beta^2}-Z^{-2}
\left(\frac{\partial Z}{\partial \beta}\right)^2\right).
\label{eqn:HeatCapa0}
\end{equation}
As shown in the Appendix~B, the formulae 
are obtained by substituting our generalized Fourtuin-Kasteleyn representation 
eq.~(\ref{eqn:GeneralizedFK}) into these equations and 
calculating the derivatives. The results are
\begin{equation}
\langle E \rangle 
=E_{\rm const}+\left\langle {\tilde E}\right\rangle_{\rm SPS},
\label{eqn:Eresult}
\end{equation}
\begin{equation}
\langle C \rangle =k_{\rm B}\beta^2\left(
\left\langle \tilde E^2 \right\rangle_{\rm SPS}-\left\langle \tilde E \right\rangle_{\rm SPS}^2 -
\left\langle \frac{\partial {\tilde E}}{\partial \beta}\right\rangle_{\rm SPS}
\right),
\label{eqn:Cresult}
\end{equation}
where
\begin{equation}
\langle {\cal O} \rangle_{\rm SPS} \equiv 
{\rm Tr}_{\{g_{ij}\},\{\vtr{S}_i\}} {\cal O}\hspace{0.5mm} P_{\rm SPS}(\{\vtr{S}_i\},\{g_{ij}\}),
\label{eqn:DefAve}
\end{equation}
\begin{equation}
E_{\rm const} = \sum\nolimits_{k<l} \Delta V_{kl}^*,
\label{eqn:DefEconst0}
\end{equation}
\begin{equation}
{\tilde E}=\sum\nolimits_{k<l}
\left\{{\tilde V}_{kl}
+\delta_{g_{kl},1}\left(\frac{\Delta V_{kl}-\Delta V_{kl}^*}
{1-P_{kl}}\right)\right\},
\label{eqn:DefEtilde}
\end{equation}
\begin{equation}
\frac{\partial {\tilde E}}{\partial \beta}
=\sum\nolimits_{k<l} \delta_{g_{kl},1}
\left(\frac{\Delta V_{kl}-\Delta V_{kl}^*}{1-P_{kl}}\right)^2P_{kl}.
\label{eqn:DefEtilde2}
\end{equation}
$P_{kl}$ in eqs.~(\ref{eqn:DefEtilde}) and (\ref{eqn:DefEtilde2}) is the 
switching probability defined by eq.~(\ref{eqn:Sprob}). 
It is quite important to note that the average in MC simulations
is equivalent to $\langle {\cal O}\rangle_{\rm SPS}$, {\it i.e.}, 
\begin{equation}
\langle {\cal O} \rangle_{\rm MC}=\langle {\cal O} \rangle_{\rm SPS}.
\end{equation}
This comes from the fact that MC simulation with the SPS algorithm 
samples states with the probability $P_{\rm SPS}(\{\vtr{S}_i\},\{g_{ij}\})$.

In the case of ${\tilde V}_{ij}=0$, $E_{\rm const}$, ${\tilde E}$, and 
$\frac{\partial {\tilde E}}{\partial \beta}$ in eqs.~(\ref{eqn:DefEconst0}), 
(\ref{eqn:DefEtilde}), and (\ref{eqn:DefEtilde2}) are reduced to the 
following forms: 
\begin{equation}
E_{\rm const}\equiv \sum\nolimits_{k<l} V_{kl}^*,
\label{eqn:DefEconst}
\end{equation}
\begin{equation}
{\tilde E}\equiv \sum\nolimits_{k<l}
\delta_{g_{kl},1}
\left(\frac{V_{kl}-V_{kl}^*}
{1-P_{kl}}\right),
\label{eqn:DefEtildeDash}
\end{equation}
\begin{equation}
\frac{\partial {\tilde E}}{\partial \beta}
=\sum\nolimits_{k<l} \delta_{g_{kl},1}
\left(\frac{V_{kl}-V_{kl}^*}{1-P_{kl}}\right)^2P_{kl},
\label{eqn:DefEtildeDash2}
\end{equation}
where $V_{ij}^*$ is a constant equal to (or greater than) 
the maximum value of $V_{ij}(\vtr{S}_i,\vtr{S}_j)$ over all 
$\vtr{S}_i$ and $\vtr{S}_j$. 
The point of the formulae is the presence of $\delta_{g_{kl},1}$ in 
${\tilde E}$ and $\frac{\partial {\tilde E}}{\partial \beta}$. 
In general, the strength of pairwise long-range interactions $V_{ij}$ decreases
with increasing the distance $r_{ij}$. Therefore, $g_{ij}$ for most of distant 
and weak interactions becomes zero (see eqs.~(\ref{eqn:Sprob}) and 
(\ref{eqn:DefGij})). This means that the computational time of 
${\tilde E}$ and its derivative is much shorter than that of 
$\sum_{k<l}V_{kl}$ which is needed in naive internal energy 
and heat capacity measurements. Although the calculation of 
the constant $E_{\rm const}$ requires ${\cal O}(N^2)$ computational time, 
it is enough to calculate it just once at the beginning of the simulation. 

In summary, the use of the formulae (eqs.~(\ref{eqn:Eresult}), (\ref{eqn:Cresult}), 
(\ref{eqn:DefEconst}), (\ref{eqn:DefEtildeDash}), and (\ref{eqn:DefEtildeDash2})) 
enables us to reduce the computational time 
for internal energy and heat capacity measurements greatly. 
This method can be used in the SCO method~\cite{SasakiMatsubara08} 
because ${\tilde V}_{ij}$ is set to be zero in the method.

\section{A formula for the replica exchange MC method}
\label{sec:formulaB}
We first explain the replica exchange MC method~\cite{HukushimaNemoto96} briefly. 
This method is quite efficient for systems with rugged energy 
landscape such as spin glasses. In this method, we consider a system 
with $M$ independent replicas and $M$ different temperatures. The $M$ 
replicas have a common Hamiltonian ${\cal H}(\vtr{S})$. The purpose of 
the method is to sample states of the $M$ replica system 
with the following equilibrium distribution
\begin{equation}
{\cal Q}_{\rm B}(\vtr{S}_1,\beta_1;\cdots;\vtr{S}_M,\beta_M)=\prod\nolimits_{m=1}^M 
P_{\rm B}(\vtr{S}_m,\beta_m),
\end{equation}
where $\vtr{S}_m$ denotes the set of real variables $\{\vtr{S}_i\}$ of 
the $m$-th replica and $P_{\rm B}$ is the Boltzmann distribution
defined by eq.~(\ref{eqn:Bdistribution}). 
In the replica exchange MC method, 
the equilibration is accelerated by exchanging 
the replica at temperature $T_m$ for that at $T_n$. 
By employing the Metropolis method, the probability of 
accepting the exchange is given as
\begin{eqnarray}
&&W_{\rm B}(\vtr{S},\beta_{m}; \vtr{S}',\beta_{n}
\rightarrow \vtr{S}',\beta_{m}; \vtr{S},\beta_{n})\nonumber\\
&&=\min \left\{ 1,\frac
{{\cal Q}_{\rm B}(\cdots;\vtr{S}',\beta_m;\cdots;\vtr{S},\beta_n;\cdots)}
{{\cal Q}_{\rm B}(\cdots;\vtr{S},\beta_m;\cdots;\vtr{S}',\beta_n;\cdots)}
\right\}\nonumber\\
&&=\min[1,X_{\rm B}],
\end{eqnarray}
where
\begin{equation}
X_{\rm B}=\exp[(\beta_m-\beta_n)({\cal H}(\vtr{S})-{\cal H}(\vtr{S}'))].
\label{eqn:DefXB}
\end{equation}

A problem of the replica exchange MC method in long-range interacting systems 
is that it costs ${\cal O}(N^2)$ computational time to calculate the 
exchange probability since $X_{\rm B}$ in eq.~(\ref{eqn:DefXB}) contains 
${\cal H}$ which consists of ${\cal O}(N^2)$ pairwise interactions. 
In order to overcome the difficulty, we consider a replica exchange MC method
whose equilibrium distribution is 
\begin{eqnarray}
&&{\cal Q}_{\rm SPS}(\vtr{S}_1,\vtr{g}_1,\beta_1;\cdots;
\vtr{S}_M,\vtr{g}_M,\beta_M)\nonumber\\
&&=\prod\nolimits_{m=1}^M P_{\rm SPS}(\vtr{S}_m,\vtr{g}_m,\beta_m),
\end{eqnarray}
where $\vtr{g}_m$ denotes the set of graph variables $\{g_{ij}\}$ of the 
$m$-th replica and $P_{\rm SPS}$ is defined by eq.~(\ref{eqn:ProbSPS}). 
It should be noted that this replica exchange MC method samples 
configuration $(\vtr{S}_1,\cdots,\vtr{S}_M)$ according to the original 
probability ${\cal Q}_{\rm B}$ since ${\cal Q}_{\rm SPS}$ is 
related to ${\cal Q}_{\rm B}$ as
\begin{eqnarray}
&&{\rm Tr}_{\vtr{g}_1,\cdots,\vtr{g}_M}{\cal Q}_{\rm SPS}
(\vtr{S}_1,\vtr{g}_1,\beta_1;\cdots;\vtr{S}_M,\vtr{g}_M,\beta_M)\nonumber\\
&&={\cal Q}_{\rm B}(\vtr{S}_1,\beta_1;\cdots;\vtr{S}_M,\beta_M),
\end{eqnarray}
where we have used eq.~(\ref{eqn:RBandSPS}). 
The probability of accepting the replica exchange is given as
\begin{eqnarray}
&&W_{\rm SPS}(\vtr{S},\vtr{g},\beta_{m}; \vtr{S}',\vtr{g}',\beta_{n}
\rightarrow \vtr{S}',\vtr{g}',\beta_{m}; \vtr{S},\vtr{g},\beta_{n})\nonumber\\
&&=\min \left\{ 1,\frac
{{\cal Q}_{\rm SPS}(\cdots;\vtr{S}',\vtr{g}',\beta_m;\cdots;
\vtr{S},\vtr{g},\beta_n;\cdots)}
{{\cal Q}_{\rm SPS}(\cdots;\vtr{S},\vtr{g},\beta_m;\cdots;
\vtr{S}',\vtr{g}',\beta_n;\cdots)}
\right\}\nonumber\\
&&=\min[1,X_{\rm SPS}],
\label{eqn:ReplicaExchange_SPS}
\end{eqnarray}
where
\begin{eqnarray}
&&\hspace*{-4mm}X_{\rm SPS}=\prod\nolimits^{(0)'} 
\exp\left[\{\beta_n-\beta_m\}
\{{\tilde V}_{ij}(\vtr{S}_i',\vtr{S}_j')+\Delta V_{ij}^*\}\right]\nonumber\\
&&\hspace*{-4mm}\times \prod\nolimits^{(0)} \exp\left[\{\beta_m-\beta_n\}
\{{\tilde V}_{ij}(\vtr{S}_i,\vtr{S}_j)+\Delta V_{ij}^*\}\right]\nonumber\\
&&\hspace*{-4mm}\times \prod\nolimits^{(1)'} \exp\left[\{\beta_n-\beta_m\}
V_{ij}(\vtr{S}_i',\vtr{S}_j')\right]
\frac{1-P_{ij}(\vtr{S}_i',\vtr{S}_j',\beta_m)}
{1-P_{ij}(\vtr{S}_i',\vtr{S}_j',\beta_n)}\nonumber\\
&&\hspace*{-4mm}\times \prod\nolimits^{(1)} \exp\left[\{\beta_m-\beta_n\}
V_{ij}(\vtr{S}_i,\vtr{S}_j)\right]
\frac{1-P_{ij}(\vtr{S}_i,\vtr{S}_j,\beta_n)}
{1-P_{ij}(\vtr{S}_i,\vtr{S}_j,\beta_m)}.\nonumber\\
\label{eqn:XSPS1st}
\end{eqnarray}
In this equation, the product $\prod^{(\alpha)}$ runs over all the pairs 
with $g_{ij}=\alpha$ and $\prod^{(\alpha)'}$ runs over those 
with $g_{ij}'=\alpha$.

When ${\tilde V}_{ij}=0$, we can rewrite the first product in the right hand side of 
eq.~(\ref{eqn:XSPS1st}) as
\begin{eqnarray}
&&\hspace*{-3mm}\prod\nolimits^{(0)'} 
\exp\left[ \{ \beta_n-\beta_m \} V_{ij}^*\right]
\nonumber\\
&&\hspace*{-3mm}=\prod_{i<j} \exp\left[ \{ \beta_n-\beta_m \} V_{ij}^*\right]
\prod\nolimits^{(1)'}
\exp\left[-\{ \beta_n-\beta_m \} V_{ij}^*\right].
\nonumber\\
\end{eqnarray}
The second product can be rewritten in a similar way. 
By substituting these results into eq.~(\ref{eqn:XSPS1st}), 
we find
\begin{eqnarray}
&&\hspace*{-8mm}X_{\rm SPS}\nonumber\\
&&\hspace*{-8mm}=\prod\nolimits^{(1)'} {\rm e}^{\left[\{\beta_n-\beta_m\}
\{V_{ij}(\vtr{S}_i',\vtr{S}_j')-V_{ij}^*\}\right]}
\frac{1-P_{ij}(\vtr{S}_i',\vtr{S}_j',\beta_m)}
{1-P_{ij}(\vtr{S}_i',\vtr{S}_j',\beta_n)}\nonumber\\
&&\hspace*{-8mm}\times \prod\nolimits^{(1)} {\rm e}^{\left[\{\beta_m-\beta_n\}
\{V_{ij}(\vtr{S}_i,\vtr{S}_j)-V_{ij}^*\}\right]}
\frac{1-P_{ij}(\vtr{S}_i,\vtr{S}_j,\beta_n)}
{1-P_{ij}(\vtr{S}_i,\vtr{S}_j,\beta_m)}.\nonumber\\
\label{eqn:XSPS2nd}
\end{eqnarray}
Since $X_{\rm SPS}$ in this formula is calculated only from the pairs 
whose graph variables are one, it can be calculated with very short computational
time as ${\tilde E}'$ and $\frac{\partial {\tilde E}'}{\partial \beta}$
in eqs.~(\ref{eqn:DefEtildeDash}) and (\ref{eqn:DefEtildeDash2}) are. 

\section{The case when long-range interactions and short-range interactions coexist}
When long-range interactions $\{ V_{ij}^{\rm (L)} \}$ and short-range interactions 
$\{ V_{ij}^{\rm (S)} \}$ 
coexist, we do not need to use the SCO method for short-range interactions
because it does not reduce the computational time significantly. In other words, 
we can switch $V_{ij}^{\rm (L)}$ to $0$ or ${\bar V}_{ij}^{\rm (L)}$ with $V_{ij}^{\rm (S)}$ 
unchanged. This can be realized by setting ${\tilde V}_{ij}^{\rm (L)}=0$ and 
${\tilde V}_{ij}^{\rm (S)}=V_{ij}^{\rm (S)}$. 
In this case, we can decompose $E_{\rm const}$, ${\tilde E}$, 
$\frac{\partial {\tilde E}}{\partial \beta}$, and $X_{\rm SPS}$ 
in eqs.~(\ref{eqn:DefEconst0}), (\ref{eqn:DefEtilde}), (\ref{eqn:DefEtilde2}), 
and (\ref{eqn:XSPS1st}) as
\begin{subequations}
\begin{align}
&E_{\rm const}=E_{\rm const}^{\rm (L)}+E_{\rm const}^{\rm (S)},\\
&{\tilde E}={\tilde E}^{\rm (L)}+{\tilde E}^{\rm (S)},\\
&\frac{\partial {\tilde E}}{\partial \beta}=
\frac{\partial {\tilde E}^{\rm (L)}}{\partial \beta}
+\frac{\partial {\tilde E}^{\rm (S)}}{\partial \beta},\\
&X_{\rm SPS}=X_{\rm SPS}^{\rm (L)}\times X_{\rm SPS}^{\rm (S)},
\end{align}
\label{eqn:LandS}
\end{subequations}
where the first terms and the second terms in the right hand sides denote
contributions from long-range interactions and those from short-range 
interactions, respectively. As we have already mentioned, the first terms 
are given by eqs.~(\ref{eqn:DefEconst}), (\ref{eqn:DefEtildeDash}), 
(\ref{eqn:DefEtildeDash2}), and (\ref{eqn:XSPS2nd}), respectively. 
On the other hand, by substituting ${\tilde V}_{kl}=V_{kl}^{\rm (S)}$ and $\Delta V_{kl}^*=0$ 
into eqs.~(\ref{eqn:DefEconst0}), (\ref{eqn:DefEtilde}), (\ref{eqn:DefEtilde2}), 
and (\ref{eqn:XSPS1st}), we readily obtain
\begin{subequations}
\begin{align}
&E_{\rm const}^{\rm (S)}=0,\\
&{\tilde E}^{\rm (S)}=-\sum\nolimits_{\langle kl \rangle} V_{kl}^{\rm (S)},\\
&\frac{\partial {\tilde E}^{\rm (S)}}{\partial \beta}=0,\\
&X_{\rm SPS}^{\rm (S)}=\prod\nolimits_{\langle kl \rangle}
\exp[(\beta_{m}-\beta_{n})(V_{kl}(\vtr{S}_k,\vtr{S}_l)-V_{kl}(\vtr{S}_k',\vtr{S}_l'))]\nonumber\\
&=X_{\rm B}^{\rm (S)},
\end{align}
\end{subequations}
where we have used the fact that all the potentials are switched to 
${\tilde V}_{kl}$. Note that $P_{kl}=1$ when ${\tilde V}_{kl}=V_{kl}$ 
and $\Delta V_{ij}^*=0$. These results are quite natural because they coincide 
with the results in the usual MC procedure. 

\begin{figure}[t]
\begin{center}
\includegraphics[width=8cm]{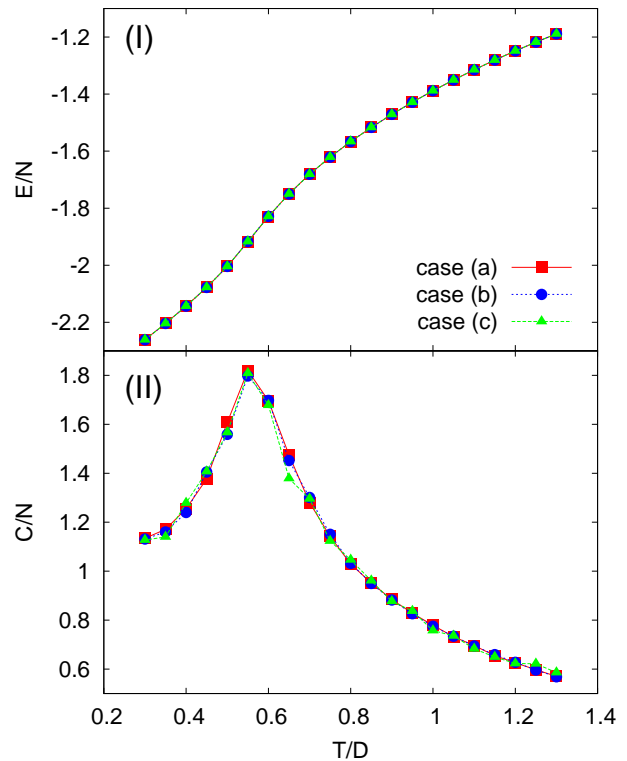}
\end{center}
\caption{(Color online) Temperature dependence of (I) internal energy $E/N$ and 
(II) heat capacity $C/N$ in the purely dipolar system. 
Simulated annealing method is used for the measurement. 
The size $N$ is $10^3$. Measurements are done in the three cases (see text).}
\label{fig:SA_EandC}
\end{figure}

\section{Numerical tests}
\label{sec:NumericalTests}
\subsection{Internal energy and heat capacity measurements}
In order to check the validity of the formulae 
for internal energy and heat capacity measurements, 
we perform MC simulations of a three dimensional magnetic dipolar system
on a $L^3$ simple cubic lattice. The boundary condition is open. 
The Hamiltonian of the system is described as
\begin{equation}
{\cal H}={\cal H}_{\rm dip}=D\sum_{i<j} \left[
\frac{\vtr{S}_i\cdot\vtr{S}_j
-3(\vtr{S}_i\cdot\vtr{r}_{ij})(\vtr{S}_j\cdot\vtr{r}_{ij})}{r_{ij}^3}\right],
\label{eqn:MagneticDipolar}
\end{equation}
where $\vtr{S}_i$ is a classical Heisenberg spin of $|\vtr{S}_{i}|=1$, $\vtr{r}_{ij}$ 
is the vector spanned from a site $i$ to $j$ in the unit of the lattice constant $a$, 
$r_{ij}=|\vtr{r}_{ij}|$, and $D=(g\mu_{\rm B}S)/a^3$. 
We hereafter call the system {\it purely dipolar system}. 

In simulations with the SCO method, 
we regard each term in the right hand side of 
eq.~(\ref{eqn:MagneticDipolar}) as $V_{ij}$. 
Since $\tilde V_{ij}=0$ in the SCO method, 
$\Delta V_{ij}$ defined by eq.~(\ref{eqn:DefofDeltaV}) is equal to $V_{ij}$. 
Interaction $V_{ij}$ has the maximum value $2D/r_{ij}^3$ when 
$\vtr{S}_i$ and $\vtr{S}_j$ are anti-parallel along $\vtr{r}_{ij}$. 
We therefore set $\Delta V_{ij}^*$, which should be equal to 
or greater than $\Delta V_{ij}$ over all $\vtr{S}_i$ and $\vtr{S}_j$, 
to $2D/r_{ij}^3$. By substituting these into eq.~(\ref{eqn:Sprob}), 
we obtain
\begin{equation}
P_{ij}=\exp\left[\beta D
 \left\{ \frac{\vtr{S}_i\cdot\vtr{S}_j
-3(\vtr{S}_i\cdot\vtr{r}_{ij})(\vtr{S}_j\cdot\vtr{r}_{ij})-2}{r_{ij}^3}
\right\}\right],
\end{equation}
Pseudointeraction ${\bar V}_{ij}$ is given by substituting $P_{ij}$ 
of the above equation into eq.~(\ref{eqn:Ppseudo}).

\begin{figure}[t]
\begin{center}
\includegraphics[height=8cm,angle=270]{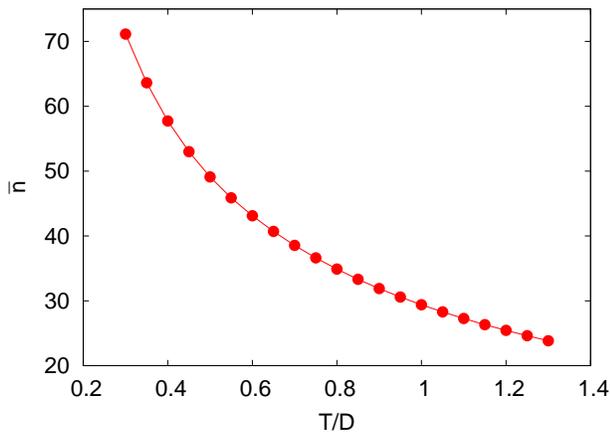}
\end{center}
\caption{(Color online) Temperature dependence of the average number $\bar n$ 
of potentials per site which are switched to ${\bar V}_{ij}$.}
\label{fig:SA_AveConnect}
\end{figure}


To make a comparison between the new measurement methods
and the usual ones, we do simulation in the following three cases:
\begin{itemize}
\item[(a)] Usual MC method with usual measurements.
\item[(b)] SCO method with usual measurements.
\item[(c)] SCO method with measurements by using 
the formulae derived in \S\ref{sec:formulaA}. 
\end{itemize}
In all the three cases, simulated annealing method is used. 
The system is gradually cooled from an initial temperature $T=1.3D$ to $0.3D$ 
in steps of $\Delta T=0.05D$. The system is kept at each temperature 
for $2\times 10^6$~MC steps. The first $10^6$~MC steps are for equilibration 
and the following $10^6$~MC steps are for measurement. The size $N$ is $10^3$. 
The size is not so large because we do simulation not only with the SCO method 
but also with a usual method which requires ${\cal O}(N^2)$ computational time 
per one MC step. 

In Fig.~\ref{fig:SA_EandC}, internal energy $E$ and heat capacity $C$ 
measured in the three cases are plotted as a function of temperature. 
We see that all the data nicely collapse into 
a single curve, showing the validity of the formulae derived in \S\ref{sec:formulaA}. 
We also see that heat capacity has a peak around $T/D=0.55$. This result is 
consistent with previous works which show the existence of a phase transition 
around this temperature~\cite{Matsushita05,Romano86}. 
Figure~\ref{fig:SA_AveConnect} shows the average number ${\bar n}$ of potentials 
per site that survive as ${\bar V}_{ij}$ in the potential switching process. 
Though ${\bar n}$ increases with decreasing temperature, ${\bar n}\approx 70$ 
even at the lowest temperature. This means that more than $90$ percent of 
interactions are cut off by being switched to ${\tilde V}_{ij}=0$. 
It is worth pointing out that the SCO method becomes 
more efficient with increasing the size. In fact, 
in the study of two-dimensional magnetic dipolar system with dipolar interactions and 
ferromagnetic exchange interactions~\cite{SasakiMatsubara08}, 
it has been found that the increase of ${\bar n}$ with size 
is very slow and ${\bar n}$ is $22.5$ even when $N=256^2=65,536$. 

\begin{figure}[t]
\begin{center}
\includegraphics[width=8cm]{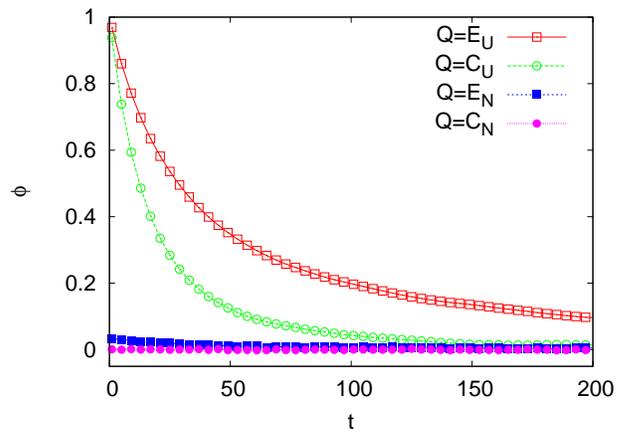}
\end{center}
\caption{(Color online) Time autocorrelation functions of the four observables 
$E_{\rm U}$, $C_{\rm U}$, $E_{\rm N}$, and $C_{\rm N}$ 
(see eq.~(\ref{eqn:4observables}) for their definitions). 
The size $N$ is $10^3$ and the temperature $T$ is $0.3D$. 
All the four observables are measured with the SCO method.}
\label{fig:SA_CorrelationTime}
\end{figure}

\subsection{Statistical error of the new measurement methods}
\label{subsec:StatisticalError}
We notice from Fig.~\ref{fig:SA_EandC} (II) that the data in the case (c), 
{\it i.e.}, those which are measured with eq.~(\ref{eqn:Cresult}), fluctuate more than 
the other data. To get some insights of this behavior, we consider estimating 
statistical error of observables. We suppose that an observable $Q$ is successively 
measured $N$ times in a MC simulation to estimate the average 
$\frac{1}{N}\sum_{\mu =1}^N Q_\mu$. We assume that the measurement is done 
every $\delta t$ steps. Although this average value is close to 
the thermal average value $\langle Q \rangle$, they are slightly different 
because the number of the measurement $N$ is finite. We hereafter call 
the difference 
$\delta Q\equiv\frac{1}{N}\sum_{\mu =1}^N Q_\mu-\langle Q \rangle$ 
{\it statistical error}. When the period 
of the measurement $N\delta t$ is much larger than the correlation time 
$\tau_{Q}$ of the observable $Q$, the expectation value of the square 
of the statistical error $\langle (\delta Q)^2 \rangle$ 
is approximately evaluated as~\cite{KrumbhaarBinder73,MonteCarloBook}
\begin{equation}
\left\langle (\delta Q)^2 \right\rangle  =\frac{1}{N}
\left[ \langle Q^2 \rangle -\langle Q \rangle^2 \right]
\left(1+2\frac{\tau_Q}{\delta t} \right).
\end{equation}
The factor $(1+2\frac{\tau_Q}{\delta t} )$ in the right hand side of the 
equation comes from the fact that $Q_{\mu}$'s measured 
successively are correlated with each other. 
From this equation, we can estimate the relative statistical error as
\begin{equation}
\frac{\sqrt{\left\langle (\delta Q)^2 \right\rangle}}{\langle Q \rangle}
= \frac{C_{\rm SE}}{\sqrt{N}},
\label{eqn:StatisticalError}
\end{equation}
where
\begin{equation}
C_{\rm SE}=\sqrt{\frac{\langle Q^2 \rangle -\langle Q \rangle^2}{\langle Q \rangle^2}
\left(1+2\frac{\tau_Q}{\delta t} \right)}. 
\label{eqn:def_of_CSE}
\end{equation}

In order to estimate $\tau_{Q}$ in eq.~(\ref{eqn:def_of_CSE}), 
we measure the normalized time autocorrelation function defined by
\begin{equation}
\phi_Q(t)\equiv\frac{\overline{Q(0)Q(t)} -\bigl(\overline{Q}\bigr)^2}
{\overline{Q^2} - \bigl(\overline{Q}\bigr)^2},
\end{equation}
where $\overline{\cdots}$ denotes the average over a sequence of the data obtained by a MC simulation. 
Since we are interested in how correlation times in internal energy and heat capacity
measurements are affected by their methods, we calculate correlation functions 
of the following four observables
\begin{subequations}
\begin{align}
& E_{\rm U}(t)={\cal H}(t),\\
& C_{\rm U}(t)=k_{\rm B}\beta^2 \Bigl({\cal H}(t)-\langle {\cal H} \rangle \Bigr)^2,\\
& E_{\rm N}(t)={\tilde E}(t)+E_{\rm const},\\
& C_{\rm N}(t)=k_{\rm B}\beta^2 \left\{
\left({\tilde E}(t)-\langle {\tilde E} \rangle\right)^2
-\frac{\partial {\tilde E}(t)}{\partial \beta}
\right\},
\end{align}
\label{eqn:4observables}
\end{subequations}
where the subscripts ${\rm U}$ and $\rm N$ denote 
the usual measurement and the new measurement by using 
the formulae derived in \S\ref{sec:formulaA}, respectively. 
The results are shown in Fig.~\ref{fig:SA_CorrelationTime}. 
All the four observables are measured with the SCO method. 
We see that correlation functions in new measurements (full symbols) 
are much smaller than those in usual measurements (open symbols). 
This probably comes from the fact that ${\tilde E}$ and 
$\frac{\partial {\tilde E}}{\partial \beta}$ can change 
without change in spin configurations since they depend on
not only $\{\vtr{S}_i\}$ but also $\{g_{ij}\}$. 
This result shows that large fluctuations in heat capacity 
measured with the new formula do not originate from 
an increase in the correlation time.

\begin{table}
\caption{Correlation time $\tau_Q$, relative variance, and $C_{\rm SE}$ of the 
four observables $E_{\rm U}$, $C_{\rm U}$, $E_{\rm N}$, and $C_{\rm N}$ 
(see eq.~(\ref{eqn:def_of_CSE}) for the definition of $C_{\rm SE}$ and 
eq.~(\ref{eqn:4observables}) for the definitions of the four observables). 
The size $N$ is $10^3$ and the temperature $T$ is $0.3D$. 
The time interval $\delta t$ in eq.~(\ref{eqn:def_of_CSE}) is 
assumed to be one in the estimation of $C_{\rm SE}$. 
}
\label{tbl:SEestimation}
\begin{ruledtabular}
\begin{tabular}{cccc}
 & $\tau_Q$ & $\displaystyle{\frac{\langle Q^2 \rangle -\langle Q \rangle^2}{\langle Q \rangle^2}}$ 
& $C_{\rm SE}$\\
\hline
$Q=E_{\rm U}$ & $96$ & $2.0\times 10^{-5}$ & $0.062$ \\
$Q=E_{\rm N}$ & $3.6$ & $5.7\times 10^{-4}$ & $0.068$ \\
$Q=C_{\rm U}$ & $25$ & $2.0$ & $10$ \\
$Q=C_{\rm N}$ & $0$ & $1.6\times 10^3$ & $40$ \\
\end{tabular}
\end{ruledtabular}
\end{table}

We next estimate the correlation time $\tau_Q$ from 
$\phi_Q(t)$ as
\begin{equation}
\tau_Q = \int_0^{\infty} {\rm d}t' \phi_Q(t').
\end{equation}
In practice, we set the lower limit to be one and adjust the upper limit 
so that it is much larger than the correlation time. For example, 
the upper limit was set to $3,000$ when we estimated 
the correlation time for $E_{\rm U}$. 
The estimated values are shown in Table~\ref{tbl:SEestimation}. 
We estimate the correlation time for $C_{\rm N}$ to be zero 
because it is negligibly small. We calculated $\tau$ for $C_{\rm N}$
with changing the upper limit from $2$ to $10,000$, and found that
it does not exceed $0.2$ for any upper limits.

In Table~\ref{tbl:SEestimation}, we also show the values of relative variance 
and those of $C_{\rm SE}$. We find that the relative variances 
in new measurements are larger than 
those in usual measurements, especially in the heat capacity. 
Concerning the internal energy, we can explicitly show 
the increase in variance from eq.~(\ref{eqn:Cresult}) as
\begin{eqnarray}
 k_{\rm B}^{-1}T^2 C=
\langle {\cal H}^2 \rangle-\langle {\cal H} \rangle^2
&=& \left\langle {\tilde E}^2 \right\rangle-\left\langle {\tilde E} \right\rangle^2
-\left\langle \frac{\partial {\tilde E}}{\partial \beta}\right\rangle\nonumber\\
&\le &\left\langle {\tilde E}^2 \right\rangle-\left\langle {\tilde E} \right\rangle^2\nonumber \\
&=& \left\langle E_{\rm N}^2 \right\rangle-\left\langle E_{\rm N} \right\rangle^2,
\label{eqn:Inequality}
\end{eqnarray}
where we have used the fact that $\frac{\partial {\tilde E}}{\partial \beta}$
defined by eq.~(\ref{eqn:DefEtilde2}) is always positive. 
Since $\langle E_{\rm N} \rangle = \langle E_{\rm U} \rangle$, 
the inequality (\ref{eqn:Inequality}) shows that 
the relative variance of $E_{\rm N}$ is larger than that of $E_{\rm U}$.
We next turn to how $C_{\rm SE}$ is affected by the measurement methods. 
We see from Table~\ref{tbl:SEestimation} that there is no significant 
difference between $C_{\rm SE}$ for $E_{\rm U}$ and that 
for $E_{\rm N}$. On the other hand, $C_{\rm SE}$ for $C_{\rm N}$ 
is four times larger than that for $C_{\rm U}$. 
This is the reason why the heat capacity in the case (c) fluctuates 
more than that in the other cases. It should be noted that the relative 
statistical error is proportional to $C_{\rm SE}$ (see eq.~(\ref{eqn:StatisticalError})). 
Equation~(\ref{eqn:StatisticalError}) also tells us that the number of
measurements with $C_{\rm N}$ should be 16 times as large as 
that with  $C_{\rm U}$ to make both the statistical errors the same.

\begin{figure}[t]
\begin{center}
\includegraphics[width=8cm]{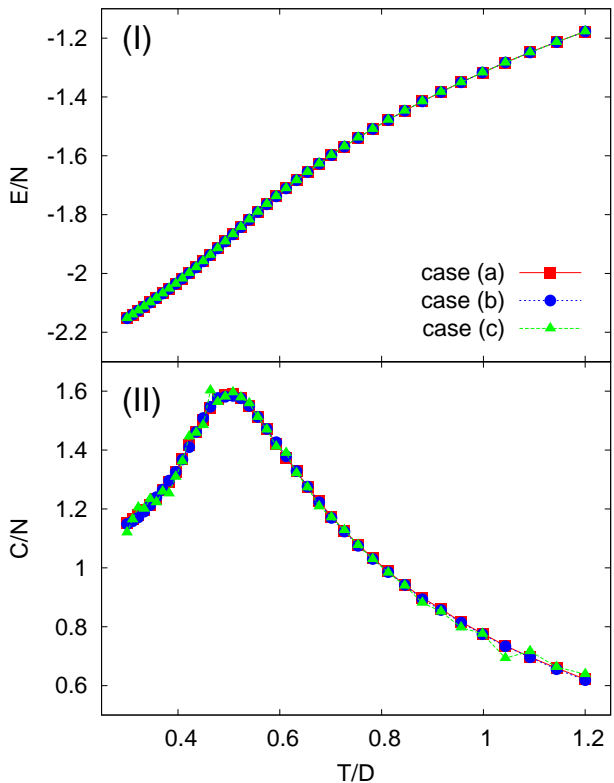}
\end{center}
\caption{(Color online) Temperature dependence of (I) internal energy $E/N$ and 
(II) heat capacity $C/N$ in the purely dipolar system. 
The replica exchange MC method is used for the measurement. 
The size $N$ is $6^3$. Measurements are done in the three cases (see text).}
\label{fig:REX_EandC}
\end{figure}

In summary, to attain a certain accuracy, 
estimation of the heat capacity with $C_{\rm N}$ requires 
larger number of measurements than that with $C_{\rm U}$ since 
fluctuations in $C_{\rm N}$ are larger than those in $C_{\rm U}$. 
On the other hand, the efficiency in the internal energy measurement 
with $E_{\rm N}$ is almost the same as that with $E_{\rm U}$ 
in the present case. When the heat capacity measurement with 
$C_{\rm N}$ does not work well, it might be possible to estimate
the heat capacity with lower statistical error by doing 
numerical differential of the internal energy which is evaluated with $E_{\rm N}$.



\begin{figure}
\begin{center}
\includegraphics[height=8cm,angle=270]{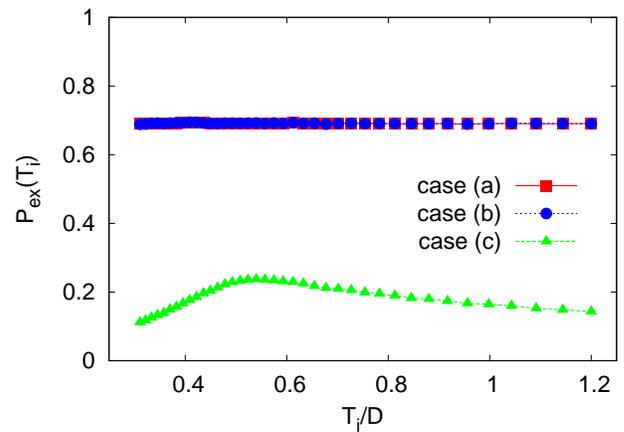}
\end{center}
\caption{(Color online) Temperature dependence of the replica exchange probability 
$P_{\rm ex}$ in the purely dipolar system. The size $N$ is $6^3$ and the number 
of temperatures is $40$.}
\label{fig:REX_ProbEX_PureDP}
\end{figure}

\subsection{Replica exchange method}
\label{subsec:REresult}
To examine the validity of the formula derived for the replica
exchange MC method, we again do 
simulations in the following three cases:
\begin{itemize}
\item[(a)] Usual MC method with usual replica exchange MC method.
\item[(b)] SCO method with usual replica exchange MC method. 
\item[(c)] SCO method with replica exchange MC method by using 
the formula derived in \S\ref{sec:formulaB}. 
\end{itemize}
In all the three cases, 
the number of MC steps for equilibration and that for measurements are $10^6$. 
The number of temperatures is $40$ and the size $N$ is $6^3$. 
Figure~\ref{fig:REX_EandC} shows the result of internal energy and heat capacity measurements. 
The measurements are done with usual method in the cases (a) and (b), and 
with our formulae (eqs.~(\ref{eqn:Eresult}) and (\ref{eqn:Cresult})) 
in the case (c). We again see that all the data nicely 
collapse into a single curve. This result clearly shows the validity 
of the formula derived in \S\ref{sec:formulaB}. 
We also see that the heat capacity in the case (c) fluctuates 
more than that in the other cases, as we have seen in Fig.~\ref{fig:SA_EandC}. 
In Fig.~\ref{fig:REX_ProbEX_PureDP}, we show the temperature dependence 
of the probability $P_{\rm ex}(T_i)$ that exchange trials between $i$-th 
and $i+1$-th replicas are accepted.
We find that $P_{\rm ex}(T_i)$ with our formula (case (c)) is smaller than 
that with usual method (the other cases). From this result, one may consider 
that our replica exchange MC method is less efficient than the usual method. 
However, this is not true because the computational time of $X_{\rm SPS}$ 
(eq.~(\ref{eqn:XSPS2nd})) is much shorter than that of $X_{\rm B}$ (eq.~(\ref{eqn:DefXB})). 

\begin{figure}
\begin{center}
\includegraphics[height=8cm,angle=270]{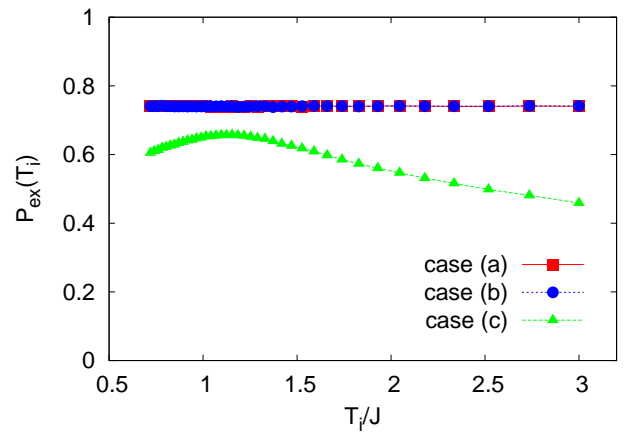}
\end{center}
\caption{(Color online) Temperature dependence of the replica exchange probability 
$P_{\rm ex}$ in the ferromagnetic dipolar system. The size $N$ is $6^3$ 
and the number of temperatures is $40$.}
\label{fig:REX_ProbEX_EXandDP}
\end{figure}

We next show results of the replica exchange MC method when long-range interactions 
and short-range interactions coexist. The Hamiltonian is consist of the long-range 
magnetic dipolar interactions (eq.~(\ref{eqn:MagneticDipolar})) 
and the short-range exchange interactions
\begin{equation}
{\cal H}_{\rm ex}=-J\sum\nolimits_{\langle ij \rangle} \vtr{S}_i\cdot\vtr{S}_j
\quad(J>0),
\label{eqn:EXinteraction}
\end{equation}
where the sum runs over the nearest neighbouring pairs. 
The ratio $D/J$ is $0.1$. We hereafter call the system {\it ferromagnetic dipolar system}. 
The SCO method is used only for the dipolar interactions. 
We again examine the three cases mentioned in the previous paragraph. 
The number of MC steps for equilibration and that for measurements are $10^6$. 
The number of temperatures is $40$ and the size $N$ is $6^3$.
We first have confirmed that internal energy and heat capacity in the three cases coincide 
with each other. We next examine the temperature dependence of 
the replica exchange probability $P_{\rm ex}(T_i)$. 
Figure~\ref{fig:REX_ProbEX_EXandDP} shows the result. We see that the reduction of 
the exchange probability in case (c) is not as large as 
that of the purely dipolar system (Fig.~\ref{fig:REX_ProbEX_PureDP}). 
This result is reasonable because 
the SCO method is used only for the long range interactions and 
its contribution ($X_{\rm SPS}^{\rm L}$ in eq.~(\ref{eqn:LandS}d)) 
is not large. Recall that the ratio $D/J$ is $0.1$. 
We also measure the ergod time defined by the average MC step 
for a specific replica to move from the lowest to the highest 
temperature and return to the lowest one. The result is shown 
in Table~\ref{tbl:ErgodTime}. In both the systems, the ergod time 
in the cases (b) and (c) is larger than than that in the case (a), 
meaning that the use of the SCO method increases the ergod time. 
However, the increase of the ergod time in the ferromagnetic dipolar system 
is not as large as that in the purely dipolar system. 
These results show that the SCO method is particularly efficient for systems 
with strong short-range interactions and weak long-range interactions. 
This feature of the SCO method has already been 
pointed out in the previous work~\cite{SasakiMatsubara08}. 


\begin{table}
\caption{Ergod time in the purely dipolar system and the ferromagnetic 
dipolar system. Measurements are done in the three cases (see text). 
}
\label{tbl:ErgodTime}
\begin{ruledtabular}
\begin{tabular}{lcc}
 & purely dipolar & ferromagnetic dipolar\\
\hline
case (a) & $1.9\times 10^4$ & $3.3\times 10^3$ \\
case (b) & $4.4\times 10^4$ & $3.5\times 10^3$ \\
case (c) & $6.2\times 10^4$ & $4.5\times 10^3$ \\
\end{tabular}
\end{ruledtabular}
\end{table}

\section{Summary}
\label{sec:summary}
In the present work, we have derived useful formulae for the SCO method 
to estimate internal energy, heat capacity, and replica exchange probability
in the replica exchange MC method. We can reduce the computational time 
of these quantities greatly by using the formulae because they only
contain terms which are not cut off by the SCO method. 
On the other hand, we have found that the use of the formulae could cause 
a decline in the efficiency of the measurement and that in the exchange probability. 
When the new methods do not work well, the analyses done in the present paper, 
such as the estimations of the statistical error, the replica exchange probability, 
and the ergod time, might be helpful to figure out the reason and to get rid of it. 
Anyway, we hope that these formulae make the SCO method 
more useful and attractive. 

The other achievement of the present work is the derivation of the new 
Fourtuin-Kasteleyn representation of the partition function, 
{\it i.e.}, eq.~(\ref{eqn:GeneralizedFK}). This representation 
is more comprehensive than the original one because our representation 
includes arbitrariness in the choice of pseudointeractions 
$\{ {\tilde V}_{ij} \}$. Furthermore, this representation can be used 
no matter whether the variables $\{\vtr{S}_i\}$ are discrete or continuous. 
We hope that this representation becomes the basis of new algorithms 
as the original Fourtuin-Kasteleyn representation lead to 
the Swendsen-Wang cluster algorithm~\cite{SwendsenWang86} 
in the Ising ferromagnetic model.

\begin{acknowledgments}
This work is supported by a Grant-in-Aid for Scientific 
Research (\#21740279) from MEXT in Japan. 
\end{acknowledgments}



\appendix
\section{Derivation of the Fourtuin-Kasteleyn representation of the 
partition function in the Ising ferromagnetic model from eq.~(\ref{eqn:GeneralizedFK})
}
In this appendix, we consider the Ising ferromagnetic model 
whose Hamiltonian is described as
\begin{equation}
{\cal H}=-J\sum\nolimits_{\langle ij \rangle}\sigma_i\sigma_j,
\end{equation}
where $\sigma_i=\pm 1$ and $J>0$. We set $V_{ij}=-J\sigma_i\sigma_j$ 
and consider a special case that ${\tilde V}_{ij}=0$ and $V_{ij}^*= J$. 
Then, we find
\begin{equation}
P_{ij}=\left\{
\begin{array}{cc}
\exp(-2\beta J) & (\sigma_i=\sigma_j),\\
1 & (\sigma_i=-\sigma_j),\\
\end{array}
\right.
\end{equation}
and
\begin{equation}
{\bar V}_{ij}=\left\{
\begin{array}{cc}
-J -\beta^{-1}\log[1-\exp(-2\beta J)],& (\sigma_i=\sigma_j),\\
\infty & (\sigma_i=-\sigma_j).\\
\end{array}
\right.
\end{equation}
By substituting this equation into eq.~(\ref{eqn:defomega}), we obtain
\begin{eqnarray}
&&\hspace*{-8mm}\omega_{ij}(\sigma_i,\sigma_j;g_{ij})\nonumber\\
&&\hspace*{-8mm}=\left\{
\begin{array}{cc}
\exp(-\beta J)& (g_{ij}=0),\vspace{1mm}\\
\exp(\beta J)[1-\exp(-2\beta J)]\delta_{\sigma_i,\sigma_j}& (g_{ij}=1).\\
\end{array}
\right.
\end{eqnarray}
It is important to note that the product $\prod_{\langle ij \rangle} \omega_{ij}$ 
is proportional to the joint distribution of $\{\vtr{S}_i\}$ and $\{g_{ij}\}$ 
introduced by Edwards and Sokal~\cite{EdowardsSokal88}. 
In this sense, we can regard the product of $\omega_{ij}$ defined by eq.~(\ref{eqn:defomega}) 
as a generalization of Edwards and Sokal's joint distribution. 

As it has been pointed out in ref.~\cite{EdowardsSokal88}, we can easily obtain 
the Fourtuin-Kasteleyn representation of the partition function from this joint distribution. 
We first rewrite $\prod_{\langle ij \rangle} \omega_{ij}$ as
\begin{eqnarray}
&&\hspace*{-17mm}\prod\nolimits_{\langle ij \rangle} \omega_{ij}={\rm e}^{N_{\rm int}\beta J}
\{\exp(-2\beta J)\}^{N_{\rm int}-N_{\rm b}}\nonumber\\
&&\hspace*{3mm}\times\{1-\exp(-2\beta J)\}^{N_{\rm b}}
\prod\nolimits_{\langle ij \rangle}^{(1)} \delta_{\sigma_i,\sigma_j},
\end{eqnarray}
where $N_{\rm int}=\sum_{\langle ij \rangle}$ and $N_{\rm b}=\sum_{\langle ij \rangle}\delta_{g_{ij},1}$. 
The product $\prod_{\langle ij \rangle}^{(1)}$ runs over all the pairs 
with $g_{ij}=1$. By using eq.~(\ref{eqn:GeneralizedFK}), we obtain
\begin{eqnarray}
&&\hspace*{-15mm}Z(\beta)={\rm e}^{N_{\rm int}\beta J} {\rm Tr}_{\{g_{ij}\}}
\{\exp(-2\beta J)\}^{N_{\rm int}-N_{\rm b}}\nonumber\\
&&\hspace*{-2mm}\times\{1-\exp(-2\beta J)\}^{N_{\rm b}}
{\rm Tr}_{\{\sigma_i\}} \prod\nolimits_{\langle ij \rangle}^{(1)} \delta_{\sigma_i,\sigma_j}.
\label{eqn:ZFK1}
\end{eqnarray}
We next consider calculating
${\rm Tr}_{\{\sigma_i\}}\prod_{\langle ij \rangle}^{(1)} \delta_{\sigma_i,\sigma_j}$
in the above equation. 
We hereafter call pairs with $g_{ij}=1$ {\it bonds} and 
a set of sites which are connected by bonds a {\it cluster}. 
Because of the presence of $\delta_{\sigma_i,\sigma_j}$, 
the values of $\sigma_i$ in a cluster are forced to be the same. 
Therefore, we find
\begin{equation}
{\rm Tr}_{\{\vtr{S}_i\}}\prod\nolimits_{\langle ij \rangle}^{(1)} \delta_{\sigma_i,\sigma_j}
= 2^{N_{\rm cluster}\{g_{ij}\}},
\end{equation}
where $N_{\rm cluster}$ is the number of clusters. By substituting this equation 
into eq.~(\ref{eqn:ZFK1}), we obtain the Fourtuin-Kasteleyn representation of the 
partition function in the Ising ferromagnetic model, {\it i.e.},
\begin{eqnarray}
&&\hspace*{-15mm}Z(\beta)={\rm e}^{N_{\rm int}\beta J} {\rm Tr}_{\{g_{ij}\}} 
\{\exp(-2\beta J)\}^{N_{\rm int}-N_{\rm b}}\nonumber\\
&&\hspace*{-2mm}\times\{1-\exp(-2\beta J)\}^{N_{\rm b}}
\times 2^{N_{\rm cluster}\{g_{ij}\}}.
\end{eqnarray}

\section{Derivation of eqs.~(\ref{eqn:Eresult})-(\ref{eqn:DefEtilde2})}
We first derive a formula for internal energy $\langle E \rangle$. 
We see from eqs.~(\ref{eqn:ProbSPS}), (\ref{eqn:GeneralizedFK}), 
and (\ref{eqn:InternalEnergy0}) that
\begin{equation}
\langle E \rangle = -Z^{-1} \frac{\partial Z}{\partial \beta}
=-\left\langle {\tilde E}'\right\rangle_{\rm SPS},
\label{eqn:InternalEnergy1}
\end{equation}
where $\langle {\cal O} \rangle_{\rm SPS}$ is defined 
by eq.~(\ref{eqn:DefAve}) and 
\begin{equation}
{\tilde E}'\equiv \sum\nolimits_{k<l}
\frac{1}{\omega_{kl}}
\frac{\partial \omega_{kl}}{\partial \beta}.
\label{eqn:Etilde1st}
\end{equation}
We have used the relation
\begin{equation}
\frac{\partial}{\partial \beta} \prod\nolimits_{i<j} \omega_{ij} 
= {\tilde E}'\prod\nolimits_{i<j}\omega_{ij} ,
\label{eqn:DevOmegaProd}
\end{equation}
to go from the second line of eq.~(\ref{eqn:InternalEnergy1}) to the third. 
The derivative of $\omega_{kl}$ in the right hand side of eq.~(\ref{eqn:Etilde1st}) 
is calculated as 
\begin{eqnarray}
&&\hspace*{-8mm}\frac{\partial \omega_{kl}}{\partial \beta}
=\frac{\partial}{\partial \beta}\left\{\delta_{g_{kl},0}
{\rm e}^{-\beta[{\tilde V}_{kl}+\Delta V_{kl}^*]}
+\delta_{g_{kl},1}
{\rm e}^{-\beta{\bar V}_{kl}}
\right\}\nonumber\\
&&\hspace*{-8mm}=-\delta_{g_{kl},0}
\left({\tilde V}_{kl}+\Delta V_{kl}^*\right)
{\rm e}^{-\beta[{\tilde V}_{kl}+\Delta V_{kl}^*]}
\nonumber\\
&&\hspace*{-6mm}\qquad 
-\delta_{g_{kl},1}
\left( {\bar V}_{kl}+
\beta\frac{\partial {\bar V}_{kl}}{\partial \beta}\right)
{\rm e}^{-\beta{\bar V}_{kl}}\nonumber\\
&&\hspace*{-8mm}=\omega_{kl}
\left\{-\delta_{g_{kl},0}\left({\tilde V}_{kl}+\Delta V_{kl}^*\right)
-\delta_{g_{kl},1}\left( {\bar V}_{kl}+\beta\frac{\partial {\bar V}_{kl}}{\partial \beta}\right)
\right\},\nonumber\\
\label{eqn:DevOmega1}
\end{eqnarray}
where we have used the identity
\begin{eqnarray}
&&\hspace*{-10mm}A_{kl}^{(0)} B_{kl}^{(0)} \delta_{g_{kl},0}
+A_{kl}^{(1)} B_{kl}^{(1)} \delta_{g_{kl},1}\nonumber\\
&&\hspace*{-10mm}=\left( A_{kl}^{(0)} \delta_{g_{kl},0}+A_{kl}^{(1)} \delta_{g_{kl},1} \right)
\left( B_{kl}^{(0)} \delta_{g_{kl},0}+B_{kl}^{(1)} \delta_{g_{kl},1} \right),\nonumber\\
\end{eqnarray}
to go from the third line of eq.~(\ref{eqn:DevOmega1}) 
to the fourth. By substituting 
\begin{equation}
\frac{\partial {\bar V}_{kl}}{\partial \beta}
=\beta^{-2}\log[1-P_{kl}]
+\beta^{-1}\frac{P_{kl}}{1-P_{kl}}
(\Delta V_{kl}-\Delta V_{kl}^*),
\end{equation}
and $\delta_{g_{kl},0}=1-\delta_{g_{kl},1}$  into eq.~(\ref{eqn:DevOmega1}), 
we obtain
\begin{equation}
\frac{\partial \omega_{kl}}{\partial \beta}
=\omega_{kl}\left\{-{\tilde V}_{kl}-\Delta V_{kl}^*
-\delta_{g_{kl},1}\left(\frac{\Delta V_{kl}-\Delta V_{kl}^*}
{1-P_{kl}}\right)\right\}.
\label{eqn:chi2}
\end{equation}
From this equation and eq.~(\ref{eqn:Etilde1st}), we find
\begin{equation}
\tilde E'=-E_{\rm const}-\tilde E,
\label{eqn:EDandE}
\end{equation}
where $E_{\rm const}$ and $\tilde E$ are defined by eqs.~(\ref{eqn:DefEconst0}) and 
(\ref{eqn:DefEtilde}), respectively. 
Equation (\ref{eqn:Eresult}) is obtained by substituting this equation 
into eq.~(\ref{eqn:InternalEnergy1}). 

We next derive a formula for heat capacity $\langle C \rangle$ from 
eq.~(\ref{eqn:HeatCapa0}). To this end, we first calculate the second 
derivative of $Z$ in the equation. 
We see from eq.~(\ref{eqn:DevOmegaProd}) that
\begin{eqnarray}
\frac{\partial^2}{\partial \beta^2} \prod\nolimits_{i<j} \omega_{ij} 
&=& \frac{\partial}{\partial \beta}{\tilde E}'\prod\nolimits_{i<j}\omega_{ij} \nonumber\\
&=& \left({\tilde E}'^2+\frac{\partial {\tilde E}'}{\partial \beta} \right)
\prod\nolimits_{i<j}\omega_{ij}.
\end{eqnarray}
By taking the trace in the both hand slides of the equation, we obtain
\begin{equation}
\frac{\partial^2 Z}{\partial \beta^2}
=Z\left\langle {\tilde E}'^2+\frac{\partial {\tilde E}'}{\partial \beta}\right\rangle_{\rm SPS} .
\end{equation}
Equation (\ref{eqn:Cresult}) is derived from this equation and eqs.~(\ref{eqn:HeatCapa0}), 
(\ref{eqn:InternalEnergy1}), and (\ref{eqn:EDandE}). 
Equation~(\ref{eqn:DefEtilde2}) is readily obtained 
from eq.~(\ref{eqn:DefEtilde}).


\begin{thebibliography}{9}
\bibitem{TimeReduction}
When pairwise interactions $V_{ij}$ of an $N$-element system decrease 
with the distance as $r_{ij}^{-\alpha}$, the number of interactions per element 
switched to ${\bar V}_{ij}$ and 
computational time per Monte Carlo step are 
${\cal O}(N)$ for $\alpha > d$, ${\cal O}(N\log N)$ for $\alpha =d$, 
and ${\cal O}(N^{2-\alpha/d})$ for $\alpha< d$, where $d$ is the spatial dimension. 
See ref.~\cite{SasakiMatsubara08} for details.
\end{thebibliography}
\end{document}